\documentclass[preprint,amssymb]{revtex4-1}
\usepackage{graphicx}
\usepackage{epstopdf}
\usepackage{bm}

\begin{document}

\title{Statistical approach to quantum mechanics \\ I: General nonrelativistic theory}
\date{\today}
\author{G. H. Goedecke}
\affiliation{Physics Department, New Mexico State University, Las Cruces, NM 88003}
\email{ggoedeck@nmsu.edu}

\begin{abstract}
In this initial paper in a series, we first discuss why classical motions of small particles should be treated statistically. 
Then we show that {\it any} attempted statistical description of {\it any} nonrelativistic classical system inevitably yields the multi-coordinate Schr\"odinger equation, with its usual boundary conditions and solutions, as an essential statistical equation for the 
system. We derive the general ``canonical quantization'' rule, that the Hamiltonian operator must be the classical Hamiltonian in 
the $N$-dimensional metric configuration space defined by the classical kinetic energy of the system, with the classical 
conjugate momentum $N$-vector replaced by $-i\hbar$ times the vector gradient operator in that space. We obtain these results 
by using conservation of probability, general tensor calculus, the Madelung transform, the Ehrenfest theorem and/or the 
Hamilton-Jacobi equation, and comparison with results for the charged harmonic oscillator in stochastic electrodynamics. We 
also provide two illustrative examples and a discussion of how coordinate trajectories could be compatible with wave properties such as interference, diffraction, and tunneling.


\end{abstract}
\maketitle

\section{Introduction}
\label{Intro}

There is no universally accepted derivation of the Schr\"odinger equation for a single spinless pointlike particle, and 
certainly no derivation of quantum mechanics as a whole. There have been many attempts to establish a classical stochastic 
foundation for the single-particle Schr\"odinger or Dirac equation, e.g., Bohm's hidden variable theory~\cite{Bohm}; the 
stochastic mechanics approach of Nelson~\cite{Nelson} and Baublitz~\cite{Baublitz}; Okamoto's approach using a complex 
Langevin 
equation~\cite{Okamoto}; Srinivasan and Sudarshan's use of quaternion measures and the Langevin equation (to obtain the Dirac 
equation)~\cite{SrinivasanSudarshan}; use of the Fokker-Planck equation~\cite{SED1}; and extensive work on the global 
statistical hidden variable theory known as stochastic electrodynamics (SED)~\cite{SED1,SED2}. Also, Gilson~\cite{Gilson} and 
Collins~\cite{Collins1,Collins2,Collins3,Collins4} used the Madelung transform~\cite{Madelung} in reverse to obtain a wave 
equation that must be obeyed by any system that satisfies a continuity equation for a non-negative density and associated 
flux in three-dimensional Eulidean space. Their wave equation has exactly the same {\em form} as the Schr\"odinger equation 
for a single point particle, but contains unknown functions instead of the potential energy and electromagnetic vector 
potential, and an unknown constant instead of Planck's constant. The unknown functions and constant would be different for, 
say, a classical fluid system than for the statistical description of a one-particle system.

The central hypothesis underlying most of the above-mentioned work is that quantum mechanics is actually a  
statistical description of the classical motions of small particles that are acted upon by both stochastic and non-stochastic force 
fields. In this 
work, we follow that hypothesis. In section~\ref{SEMP} we first mention the features and failures of classical electrodynamics 
that necessitate a statistical description of the motions of small charged particles.  Then we provide a concise derivation of 
Collins' statistical wave equation for a single electric monopole particle. Without reference to the Schr\"odinger equation, we show that the unknown quantities in the statistical 
equation must be 
the particle's potential energy, the electromagnetic vector potential, and Planck's constant. Also we note that the wavefuction of the equation has the same significance and must satisfy the same boundary conditions as the Schr\"odinger wavefunction, whereby the statistical wave equation and its solutions are indeed 
identical in all cases to the axiomatic Schr\"odinger equation and its solutions for such a particle.  In section~\ref{schrodinger} we develop the 
mathematical formalism 
for the statistical description of the nonrelativistic motions of one or many particles, each of which may have mass, electric 
charge, spin and associated magnetic dipole moment, and possibly other properties, such that $N$ generalized coordinates are needed to describe the system classically. We begin with a generic nonrelativistic classical 
Lagrangian and corresponding Hamiltonian for $N$ generalized curvilinear coordinates, identify the metric of this $N$-space, 
and obtain the generalized Newton's second law. 
Then, because the system must be treated statistically for the same reasons discussed in section~\ref{SEMP}, we write down the ensemble-averaged
continuity equation for the generalized-coordinate probability density and flux, and show by the same 
methods used in section~\ref{SEMP} that this $N$-space continuity equation indeed implies the usual $N$-space Schr\"odinger 
equation involving the expected general canonical quantization and the usual boundary conditions, solutions, and significance for the N-space wavefunction. In section~\ref{examples}, we provide analyses for two important 
examples. The first example is a 
rudimentary two-particle atom, which yields the usual hydrogenic Schr\"odinger equation for spinless particles. The second 
example is a system of arbitrarily many spinless pointlike identical particles, which yields the expected nonrelativistic many-body quantum field theory for spinless bosons or fermions as a statistical theory.  In section~\ref{summary}, we provide a summary 
and discussion of our results, including possible differences in interpretation between the statistical nonrelativistic quantum mechanics developed herein and the conventional axiomatic nonrelativistic quantum mechanics; a preview of the next paper on particles with spin and 
magnetic moment; and 
plans and suggestions for future work. In the Appendix, we provide a summary of the general tensor calculus methods used in 
section~\ref{schrodinger}. 

\section{Single electric monopole particle}
\label{SEMP}

In subsection~\ref{failures} below, we discuss the reasons why classical electrodynamics generally cannot be used to obtain the 
detailed motions of a pointlike electric monopole particle of small mass. In subsection~\ref{statistics}, we show that the correct 
statistical description of the nonrelativistic motions of such a particle is indeed the conventional Schr\"odinger equation in 
three spatial dimensions along with its usual boundary conditions and solutions. 

\subsection{Failures of classical electrodynamics}
\label{failures}

In order to appreciate the two principal features of classical electrodynamics (CED) that make it unable to predict the 
detailed motions of very small particles, one may consider a very simple system consisting of one indestructible particle 
having only the attributes of electric charge $q$, mass $m$, and center-of-mass (CM) position vector $\bm{X}(t)$ as a 
function of time $t$, acted upon by electromagnetic fields. For this system, nonrelativistic CED consists of the Maxwell equations and Newton's second 
law for $\bm{X}(t)$, with Lorentz force containing both arbitrary external electromagnetic fields and appropriate self-fields. The self-fields and thus a radiation 
reaction force should be present; otherwise, e.g., Newton's law would predict that a classical orbit in an isolated hydrogen 
atom would be stable despite its energy loss due to radiation.

The first major failure of CED is that it offers no satisfactory representation of the radiation reaction force that acts on 
an accelerated charged particle. All attempts to derive radiation reaction self-forces from causal self-fields yield either 
runaway or acausal solutions of Newton's 2nd law, as well as unbounded self-energy, in the point particle limit. The unbounded 
self-energy is not a fatal problem because it can be absorbed into a renormalized mass. One can avoid both the unbounded self-
energy and runaway/acausal pathologies by using an extended-particle model, but a mass renormalization, albeit finite, is 
still needed, and the effective Newton's law contains time derivatives of $\bm{X}(t)$ of arbitrarily high order, or becomes 
(approximately) a differential-difference equation~\cite{Goedecke1975} that is dependent on the structure of the model 
particle.  Special relativistic treatments such as the Lorentz-Dirac equation do not remove the difficulties. 

The second and even more important major failure of CED stems from the conventional interpretation of the classical Maxwell equations that the external electromagnetic 
fields acting on any particle are the sum of presumably causal fields emitted by other particles/sources. In principle, these external field sources are all the other particles in the universe. These causal fields include a broadband radiation field that cannot be described precisely unless the detailed motions of its sources are known exactly, which is manifestly impossible. So this radiation field  
should be expressed as a stochastic field. We shall call this field the ``universal background field'' (UBF), 
and later introduce a specific statistical model for it. 

If an object is sufficiently massive (macroscopic), neither the UBF 
nor the radiation reaction self-fields acting on its constituent microscopic particles should have measurable effect on the CM or rigid rotational motion of the object, so {\it e.g.} the CM 
coordinate of such an object should satisfy the classical Newton's 2nd law in the presence of known applied force fields. However, any particle/object that interacts electromagnetically 
in molecular, atomic and nuclear processes has such small mass that even a fairly weak UBF should cause its CM 
position $\bm{X}(t)$, its overall angular velocity $\bm{\omega}(t)$ about its CM, and possibly other (internal) coordinates to 
execute considerable ``zitterbewegung'', rapidly oscillating/fluctuating motion, which in turn should produce significant 
radiation reaction. Even if one could find closed-form expressions for $\bm{X}(t)$ and $\bm{\omega}(t)$ by solving the classical 
motion equations for such a particle, which is almost never possible, the expressions would contain the stochastic variables 
carried by the UBF. The conclusion is inevitable: The classical motions of small-mass particles should be described statistically.

\subsection {Statistics of single-particle translational motion}
\label{statistics}

\subsubsection{Position probability densities and continuity equations}
\label{3D continuity}
In {\it any} statistical description of the motion of a pointlike particle that has the attributes of 
electric charge $q$, inertial mass $m$, and CM position $\bm{X}(t)$, but no rotational or other degrees of freedom, one fundamental quantity is the 
position probability density $\rho(x,t)$, such that $\rho(x,t)d^3x$ is the probability that the particle CM is in volume 
element $d^3x$ at location $x$ at time $t$. In this section, $x$ stands for the three independent Cartesian coordinate 
variables $x=x^1,x^2,x^3$ in a Euclidean 3-space, and $X(t)=X^1(t),X^2(t),X^3(t)$ for the three Cartesian coordinates of the 
particle CM as functions of $t$. The CM moves on a trajectory given by $x^i=X^i(t),\;i = 1,2,3$. The CM position vector may be written
$\bm{X}(t)=\bm{\hat{e}}_i X^i(t)$, where the three Cartesian basis vectors $\bm{\hat{e}}_i = \bm{\hat{e}}^i$ are the 
orthogonal unit vectors of right-handed Cartesian coordinates. Repeated coordinate indices in any expression are summed over, from 1 to 3 
in this case. We indicate the coordinates by superscripts in accordance with the standard notation of general tensor calculus 
(see the Appendix) that we must use in section~\ref{schrodinger} below. The fine-grained CM position probability density is  
$$\rho^f(x,t)=\prod_{i=1}^3 \delta \left( x^i-X^i(t) \right),$$
where $\delta$ is the Dirac delta. The associated fine-grained probability current density (flux) is  
$$\bm{j}^f(x,t)=\bm{\hat{e}}_i \dot{X}^i(t)\rho^f = \bm{\dot{X}}(t)\rho^f.$$ 
Note that these quantities satisfy the continuity relation
\begin{equation}
\label{eq1}
\partial_t \rho^f + \bm{\nabla}\bm{\cdot}\bm{j}^f = 0,
\end{equation}
which guarantees conservation of probability. Here, the gradient operator $\bm{\nabla}=\bm{\hat{e}}^j \partial_j$ and
$\partial_j = \partial/\partial x^j$, $\partial_t = \partial/\partial t$.
However, these fine-grained densities are rarely useful, because in virtually no cases can we obtain expressions or accurate 
numerical solutions for $\bm{X}(t)$ for small masses, as mentioned above.

     A statistical description involving “smooth” densities is needed, so some kind of averaging of the fine-grained densities must be done. As we shall see below, the average that should be used is the ensemble average over the very many (perhaps infinitely many) stochastic variables in the UBF. Then the primary quantities of interest are the ensemble averages of the fine-grained densities, $\rho = \left<\rho^f \right>$  and $\bm{j} = \left<\bm{j}^f \right>$ , where $\left<\;\right>$ signifies the ensemble average. Clearly, the averaged densities also satisfy the continuity equation         
\begin{equation}
\label{eq2}
\partial_t \rho + \bm{\nabla}\bm{\cdot}\bm{j} = 0,
\end{equation}
since the processes of ensemble averaging and spacetime derivation commute. Furthermore, these smooth densities must satisfy eq.~(\ref{eq2}) regardless of whether detailed particle trajectories and fine-grained densities even exist, simply because probability must be conserved.  Also we note that eq.~(\ref{eq2}) must be satisfied irrespective of what stochastic process is considered, {\it e.g.}, Markovian or not, and independently of what kind of stochastic dynamics is considered, {\it e.g.}, the Langevin equation, the Fokker-Planck equation, etc., and independently of what kind of position-velocity or position-momentum phase space treatment may be valid. In particular, the position probability density, fine-grained or smooth, is always related to the position-velocity ($x,v$) phase space probability density $f(x,v,t)$ by $\rho(x,t) = \int d^3v f(x,v,t)$, where the integral is over all velocity space; and similarly for position-momentum phase space. To the best of our knowledge, all statistical treatments of any kind have (usually tacitly) assumed that the densities are smooth differentiable functions, {\it e.g.}, not containing Dirac deltas. We began above with fine-grained position probability densities, analogously to the Klimontovich phase space approach in plasma physics, in order to emphasize that the statistical Schr\"odinger equation obtained in what follows seems to be compatible with coordinate trajectories.

\subsubsection{The Schr\"odinger equation}
\label{3DSEQ}
     In this subsection, we show that the smooth-density continuity equation plus a very few other requirements {\it inevitably} yield the conventional Schrödinger equation (SEQ) as a correct statistical description of the single electric monopole particle system treated above. The first part of the derivation was published in 1977 by R.E.Collins~\cite{Collins1}. We provide a concise form of Collins' derivation below not only for clarity but also because we can use the same set of equations in sec.~\ref{schrodinger}. The first step is to define a probability flow velocity field $\bm{v}(x,t)$ by writing
\begin{equation}
\label{eq3}
\bm{j}(x,t)=\rho(x,t)\bm{v}(x,t),
\end{equation}
so that $\bm{v}$ is analogous to a fluid flow velocity field. The physical significance of $\bm{v}$ will become apparent in the next subsection. It will be smooth since both $\rho$ and $\bm{j}$ are smooth. This definition may always be made provided that $\bm{j} = 0$ if $\rho = 0$, which is the case in any physical theory. With no loss of generality, one may express $\bm{v}$ as the sum of a gradient and another vector field that is not a gradient:
\begin{equation}
\label{eq4}
\bm{v}(x,t) = \frac{\Gamma}{m}\bm{\nabla}\Phi(x,t) - \bm{u}(x,t)
\end{equation}
where $\Phi(x,t)$ is real-valued and required to be dimensionless, so that $\Gamma$ is an unknown real constant that has the dimension of angular momentum, and $\bm{u}(x,t)$ is an unknown real-valued vector field, not a gradient, having dimension velocity. That field may be required to have zero divergence, in which case eq.~(\ref{eq4}) is the Helmholtz theorem. The next step is to define the complex-valued function $\psi(x,t)$ as
\begin{equation}
\label{eq5}     
\psi(x,t)=\sqrt{\rho}\,\exp(i\Phi)
\end{equation}
This relation is known as the Madelung transform~\cite{Madelung} when applied to the usual Schr\"odinger equation; here, it is being used in reverse. Note that $\rho\geq{0}$, so $\sqrt{\rho}$ is well-defined and real $\geq{0}$ if $\rho$ is smooth, e.g., not a product of Dirac deltas as is the fine-grained density. Equations~(\ref{eq3}-\ref{eq5}) combine to yield
\begin{equation}
\label{eq6}     
\rho=\psi^*\psi
\end{equation}
\begin{equation}
\label{eq7}     
\bm{j}=\frac{\Gamma}{2im}(\psi^*\bm{\nabla}\psi - \psi\bm{\nabla}\psi^*) - \bm{u}\psi^*\psi
\end{equation}
Then, requiring that the continuity equation be satisfied yields easily
\begin{equation}
\label{eq8}     
\frac{O\psi}{\psi} = \frac{O^*\psi^*}{\psi^*} = W(x,t),
\end{equation}
where $O$ is the operator
\begin{equation}
\label{eq9}     
O = i\Gamma\partial_t - \frac{1}{2m}(-i\Gamma\bm{\nabla} - m\bm{u})^2
\end{equation}
and $W(x,t)$ is an unknown real-valued scalar field having dimension energy. Therefore, the equation that must be satisfied by $\psi$ is
\begin{equation}
\label{eq10}     
 i\Gamma\partial_t\psi = \frac{1}{2m}(-i\Gamma\bm{\nabla} - m\bm{u})^2\psi + W\psi.
\end{equation}
This statistical wave equation, which has exactly the same {\it form} as the conventional SEQ for this system, was the culmination of Collins' purely mathematical derivation~\cite{Collins1}; Gilson's  earlier work~\cite{Gilson} did not include the vector field $\bm{u}$. Note that the meanings of $\rho$ and $\bm{j}$ were established {\it a priori} (which was not the case during the original development and interpretation of the SEQ), whereby $\psi$  must be bounded, single-valued, first-order differentiable except at (unphysical) Dirac delta potentials, and magnitude square integrable over all space, the same conditions that are imposed on the conventional SEQ wavefunction for this system. (A recent paper~\cite{Goedecke2010} discusses several aspects of the inverse Madelung transform method, including why nodal surfaces of bound states do not exist except at spatial infinity, where $\psi\rightarrow 0$ rapidly enough that $\bm{\nabla}\psi$ is also zero there, whereby $\bm{\nabla}\psi$ exists everywhere except at Dirac delta potentials.) However, the presence of unknown functions $W$ and $\bm{u}$ and an unknown constant $\Gamma$ in eq.~(\ref{eq10}) might well cause one to regard Collins' result as less than a full derivation of the SEQ. The doubt is strengthened when one realizes that eq.~(\ref{eq10}) must apply for {\it any} smooth quantities  $\rho$  and $\bm{j}$ that satisfy a continuity relation, such as the smoothed mass density and mass flux in a classical fluid. For such an application, the unknown functions and constant must in general be quite different than in the statistical wave equation for a single point particle; the smoothing process may be different; and the interpretation of the wave equation must be different, as discussed by several authors~\cite{Collins1,Goedecke2010,Klein}.

     Of course, one can identify $W$, $\bm{u}$, and $\Gamma$ in eq.~(\ref{eq10}) by comparison of its predictions with experimental results, or equivalently by comparison with the known SEQ, but doing so keeps the derivation purely mathematical and does not appear to resolve the objections noted above. So we ask: Are there any ways to identify these quantities {\it a priori} using only general physical theory? The answer is yes. For example, Collins~\cite{Collins1} showed that the expectation energy of the particle is given by the usual quantum expression only if $W$, $\bm{u}$, and $\Gamma$ are the expected choices. Several authors, e.g. de la Pe\~{n}a and Cetto~\cite{SED1}, have provided arguments to identify $W$ and $\bm{u}$.  In the remainder of this subsection, we provide physical rationale to identify all three of these quantities. 
     
    {\bf Identification of unknown functions.} One straightforward way to identify $W$ and $\bm{u}$ is to apply the Ehrenfest theorem~\cite{Schiff} in reverse. That is, on the basis of the known statistical significance of $\rho$ and $\bm{j}$, we must require that the usual nonrelativistic Newton's second law be valid in the mean. We begin as usual by defining the statistical mean or expectation CM position vector as $\overline{\bm{X}}(t) = \int d^3x \bm{x}\rho$. Then we obtain the mean CM velocity as $\bm{\overline{V}}(t) = d_t\overline{\bm{X}}(t) = \int d^3x \bm{x}\partial_t\rho$. Then we use the continuity equation and an integration by parts, assuming surface integrals vanish at spatial infinity, to obtain
\begin{equation}
\label{eq11}     
\bm{\overline{V}}(t) = \int d^{3}x\rho\bm{v} = \frac{1}{m}\int  d^{3}x \psi^*(-i\Gamma\bm{\nabla} - m\bm{u})\psi,
\end{equation}
where the second equality results from using eq.~(\ref{eq7}) and another integration by parts, again assuming that surface integrals vanish. [Note that we could have begun with a definition of the mean velocity, instead of the mean position vector used in many textbooks. We must do the former in the generalized curvilinear coordinates that we need in section~\ref{schrodinger}; the two starting points are equivalent in Cartesian coordinates in the Euclidean 3-space considered here.] Also, note that eq. (\ref{eq11}) provides a physical significance for the vector field $\bm{v}$, such that $\bm{j} = \rho\bm{v}$ is the probability flux density that yields the mean particle velocity $\bm{\overline{V}}(t)$.

Now, take the time derivative of eq.~(\ref{eq11}), using eq.~(\ref{eq10}). It is straightforward to show that the result is
\begin{equation}
\label{eq12}     
\frac{d\bm{\overline{V}}}{dt} = \frac{1}{m}\int d^{3}x \psi^*\bm{F}^{op}\psi\,
\end{equation}
where
\begin{equation}
\label{eq13}     
	\bm{F}^{op} = -\bm{\nabla}W - m\partial_t\bm{u} + \mbox{\small$\frac{1}{2}$}m[\bm{v}^{op}\times{(\bm{\nabla}\times{\bm{u}})} - (\bm{\nabla}\times{\bm{u}})\times{\bm{v}^{op}}],
\end{equation}
with
\begin{equation}
\label{eq14}     
\bm{v}^{op} = m^{-1}\bm{p}^{op} - \bm{u};\;\;\;\;\bm{p}^{op} = -i\Gamma\bm{\nabla}.
\end{equation}
Just as for eq.~(\ref{eq11}), the validity of eq.~(\ref{eq12}) depends on requiring that surface integrals at spatial infinity vanish, often accomplished by requiring the integrands to satisfy periodic boundary conditions on the surfaces of a cubical box of side length $L\rightarrow\infty$.   These requirements are equivalent to assuming that all the operators indicated by the superscript ``op" are Hermitian. (Note that these same requirements must also be applied in the usual treatment of the axiomatic SEQ.) From eqs.~(\ref{eq12}) - (\ref{eq14}), it is evident that Newton's second law is valid in the mean if and only if
\begin{equation}
\label{eq15}     
W = q\varphi + W_{0};\;\;\;\;    \bm{u}=q\bm{A}/mc,
\end{equation}
where $\varphi$  and $\bm{A}$ are electromagnetic scalar and vector potentials, respectively, and $W_0$ is some other potential energy (say, gravitational) that can affect this monopole particle. This conclusion follows because only with these identifications do we obtain the correct (Hermitian) Lorentz force operator from eq.~(\ref{eq13}),
\begin{equation}
\label{eq16}     
\bm{F}^{op} = -\bm{\nabla}W_0 + q[\bm{E}+(2c)^{-1}(\bm{v}^{op}\times{\bm{B}} - \bm{B}\times{\bm{v}^{op}})],
\end{equation}
where $\bm{E} =  -\bm{\nabla}\varphi - c^{-1}\partial_t\bm{A}$  is the electric field, and $\bm{B} = \bm{\nabla}\times{\bm{A}}$  is the magnetic field (flux density), in Gaussian units. After making the identifications in eq.~(\ref{eq15}), we may rewrite eq.~(\ref{eq10}) in the form
\begin{equation}
\label{eq17}     
i\Gamma\partial_t\psi = H^{op}\psi,
\end{equation}
where
\begin{equation}
\label{eq18}     
H^{op} =\frac{1}{2m}(\bm{p}^{op} - \frac{q}{c}\bm{A})^2 + q\varphi + W_0
\end{equation}
has the form of the classical nonrelativistic Hamiltonian of the system, with conjugate momentum $\bm{P}$ replaced by $\bm{p}^{op}$, including whatever fields $\bm{A}$, $\varphi$, and $W_0$ were appropriate for the classical Hamiltonian. Note that eqs.~(\ref{eq17}) and~(\ref{eq18}) also seem to apply to an electrically neutral particle, as conventionally assumed.

     There is another presumably equivalent but more direct method that can be used to identify the unknown functions. As is easily shown, {\it e.g.} see Goedecke and Davis~\cite{Goedecke2010}, substitution of eq.~(\ref{eq5}) into eq.~(\ref{eq10}) yields two equations that must be satisfied. One is the probability continuity equation itself, while the other is
     
$\Gamma\partial_t\Phi + (2m)^{-1}(\Gamma\bm{\nabla}\Phi - m\bm{u})^2 + W - (\Gamma^2/2m)(\nabla^2\rho^{1/2}/\rho^{1/2}) = 0.$  

The last term on the left-hand side is proportional to the so-called “quantum-mechanical potential (energy)”. If it is negligible, then this equation must reduce to the classical Hamilton-Jacobi equation for Hamilton's principal function $S=\Gamma\Phi$ for the electric monopole system, which occurs if and only if eq.~(\ref{eq15}) is satisfied.

At this point, we have shown something that seems quite remarkable, namely, that if {\it for any reason} we choose to describe the classical motion of an electric monopole or uncharged pointlike particle statistically, then a correct statistical description is the statistical Schr\"odinger equation ~(\ref{eq17}), which is indistinguishable from the conventional equation, and in which the wavefunction has the conventional meaning and satisfies exactly the usual conditions. Therefore, for any given quantized or c-number fields $\varphi, \bm{A}$, and $W_0$, {\it all} solutions of the statistical 
Schr\"odinger equation are exactly the same as the conventional ones, except that $\hbar$ is replaced by an unknown constant that we denoted by $\Gamma$.
 
     {\bf Identification of unknown constant.} It is apparent that one must choose $\Gamma = \hbar$ in order that eq.~(\ref{eq17}) be formally identical to the nonrelativistic SEQ for the electric monopole particle system. We seek a physical rationale for that choice. We expect that the value of $\Gamma$ is determined by the universal background field (UBF) mentioned earlier, because that is the field that compels a statistical description. Therefore we require that the value of $\Gamma$ be universal, the same for all systems in the universe (or at least in a fairly extensive region of the universe). This requirement rules out the possibility that the UBF fields are just thermal fields; therefore, the mean energy density of the UBF must be considerably greater than that of all thermal fields, even those in stellar atmospheres. So we seek an omnipresent temperature-independent high-energy-density stochastic electromagnetic field as the overwhelmingly dominant part of the UBF. 

     Such a field has been proposed and investigated extensively. Beginning in the early 1960's, many authors~\cite{SED1, SED2} contributed to the development of what became known as “stochastic electrodynamics” (SED). The central thesis of SED is that all particles are acted upon by a universal stochastic background radiation field, omnipresent even at the absolute zero of temperature, which in turn causes particles that interact electromagnetically to perform “zero-point oscillations” and via radiation reaction be in dynamical equilibrium with the background field, in the absence of other external fields. This hypothesized field became known as the (stochastic) zero-point field (SZPF). The SZPF was modeled as a classical but stochastic free electromagnetic field, not referred to its sources, having expectation energy $(\hbar\omega_k/2,\;\omega_k = kc)$ per plane wave normal mode with propagation vector $\bm{k}$ and transverse polarization index 1 or 2, which is the same as the energy eigenvalue of a quantum vacuum field transverse normal mode. In 1969, Boyer~\cite{Boyer} showed in detail that the presence of the SZPF plus a classical stochastic statistically independent thermal field yields exactly the Planck blackbody radiation spectrum for a collection of oscillators, without quantization of oscillator levels. During the development of SED, many authors also showed that the classical nonrelativistic Newton’s 2nd law applied to the system of a charged-particle isotropic harmonic oscillator with natural angular frequency $\omega_0$, and with damping force given by an Abraham-Lorentz radiation reaction term, acted upon by the SZPF in electric dipole approximation, yields the ensemble-average energy $3\hbar\omega_0/2$, the same as the quantum ground state energy. Some of these authors went further, deriving the SEQ and in some cases its closed-form solutions from the ensemble average of the fine-grained position-momentum phase space distribution function for the oscillator acted upon by the SZPF and other specified non-random radiation fields. These results showed that the appropriate smoothing average of the fine-grained probability density and flux is indeed the ensemble average, as mentioned just before eq.~(\ref{eq2}). For example, Goedecke~\cite{Goedecke1983} showed that the usual quantum electrodynamics results for transition probabilities per unit time for resonance absorption, stimulated emission, and spontaneous emission were predicted, with the spontaneous emission occuring automatically, without the triggering needed in the Crisp-Jaynes-Stroud semiclassical theory~\cite{CrispJaynes}, and without quantization of electromagnetic fields. These SED derivations of a statistical SEQ, and of a companion equation that restricts initial conditions on the wavefunction so that the resulting Wigner phase space distribution can never be negative, are valid only for the nonrelativistic harmonic oscillator system in electric dipole approximation. They are completely different from the derivation of the statistical SEQ presented herein, which as shown above applies to all nonrelativistic single-particle electric monopole systems, with no restriction to the electric dipole approximation for any applied electromagnetic fields.    

For our purposes, the important aspects of the discussion above are that i) the SZPF contains Planck's constant, and the SED results for a charged harmonic oscillator immersed in the SZPF imply that the (presumed universal) constant $\Gamma$ that appears in some of the equations~(\ref{eq10}) -~(\ref{eq14}) must be $\hbar$; and ii) the correct smoothing average of the fine-grained position probability density is the ensemble average over the stochastic variables in the SZPF, whereby the statistical Schr\"odinger equation derived above is itself an ensemble-averaged equation. We will discuss this feature in section~\ref{summary}.

At this point, then, we have shown that if we attempt {\it any} statistical description of the classical nonrelativistic CM motion of a pointlike electric monopole particle immersed in the stochastic zero-point field and other force fields, then a correct ensemble-averaged statistical equation is the conventional Schr\"odinger equation in which the wavefunction has the conventional meaning and satisfies exactly the same conditions, so that all solutions are exactly the same as the conventional ones, for {\it any} choice of the potentials $\varphi, \bm{A}$, and $W_0$.
 
\section{General Schr\"odinger equation}
\label{schrodinger}
In this section we consider nonrelativistic systems that require $N$ generalized curvilinear coordinates to describe. For example, for a classical nonrelativistic system of $N_p$ identical particles with spin, the coordinates for each particle could be three Cartesian coordinates (or three spherical polar coordinates or ...) for the CM motion, and three Euler angles for the rotational motion, or altogether $N = 6N_p$ generalized coordinates that comprise the ``configuration space" of the system. [We treat spin in detail in the next paper in the series]. In subsection~\ref{genCM}, we discuss the forms of the classical nonrelativistic Lagrangian, Hamiltonian, and coordinate motion equations for virtually all systems that require $N$ coordinates, identifying the metric and the covariant and contravariant basis vectors and their crucial properties in the $N$-space. In subsection~\ref{genSEQ} we begin with the generalized-coordinate probability continuity equation, and show that the same approach used for an electric monopole in section~\ref{SEMP} yields the $N$-space Schr\"odinger equation, including the generalized conjugate momentum operators and Hamiltonian operator.

\subsection{Classical mechanics in generalized coordinates}
\label{genCM}
     Although the tradition in nonrelativistic classical mechanics is to write generalized coordinates as $q_i$, with subscript indices, here we employ superscripts, and $x^i$ instead of $q_i$, in order to take advantage of the conventional general tensor calculus description of coordinate manifolds as metric spaces. So we represent the $N$ generalized coordinates needed for a system under consideration by the set $x = [x^1, \ldots, x^N]$, where each $x^p$ is an independent real continuous variable that may have any dimension and any range. A classical system moves on a trajectory in this $N$-space given by $[x^p = X^p(t),\;p = (1, \ldots, N)]$.

     The nonrelativistic Newton's second law for the coordinates of a system always seems to result in coupled second order differential equations that are linear in the coordinate accelerations, at most quadratic in the coordinate velocities, and linear in the electromagnetic, gravitational, or other possible fields acting on the particles. (A discussion of why this should be the case is best left to the relativistic treatment to follow in a later paper.) Therefore, in this work, we consider only the most general form of Lagrangian that will yield such motion equations for the coordinates. The machinery of general tensor calculus in $N$ dimensions (see the Appendix) allows us to write down that generic Lagrangian in a familiar form. First, we define the $N$-velocity vector $\bm{V}(t)$ in terms of its contravariant components $\dot{X}^p$ and the covariant basis vectors $\bm{e}_p(X)$:
\begin{equation}
\label{eq19}     
\bm{V} = \bm{e}_p(X)\dot{X}^p.
\end{equation}
Then, the most general appropriate Lagrangian having the dimension of energy is
\begin{equation}
\label{eq20}     
L = \mbox{\small$\frac{1}{2}$}m\bm{V}\bm{\cdot}\bm{V} + m\bm{u}(X,t)\bm{\cdot}\bm{V} - W(X,t),
\end{equation}
where $\bm{u}(X,t)$ is an $N$-vector field having dimension velocity, $W$ is an $N$-scalar field having dimension energy, $m$  is a parameter having dimension mass, and the dot $\bm{\cdot}$ indicates the generalized dot or scalar product, as discussed in the Appendix. Note that $\bm{u}$ and $W$ may have both explicit and implicit time-dependence. Clearly, the first term in $L$ is a generalized kinetic energy. If all the particles in the system have the same mass, then one could choose $m$ to be that mass; in general, $m$ is any convenient constant having dimension mass, and actual masses or other appropriate parameters will be contained in the basis vectors and thus in the metric. If we write $\bm{u}$  as a linear combination of the contravariant basis vectors, as we may for any vector field, and use eq.~(\ref{eq19}) and eq.~(\ref{eqA1}), we find an equivalent expression for $L$,
\begin{equation}
\label{eq21}     
L = \mbox{\small$\frac{1}{2}$}mg_{pq}(X)\dot{X}^p\dot{X}^q + mu_p(X,t)\dot{X}^p - W(X,t),
\end{equation}
where $g_{pq}(X) = \bm{e}_p(X)\bm{\cdot}\bm{e}_q(X)$ is defined as the (covariant) metric (see the Appendix). Note that $g_{pq}$ is indeed the conventional metric as defined by eq.~(\ref{eqA7}). The covariant components of the conjugate momentum vector are given by
\begin{equation}
\label{eq22}     
P_s = \partial L/\partial\dot{X}^s =m(g_{sq}\dot{X}^q + u_s).
\end{equation}
In boldface N-vector notation this equation is simply
\begin{equation}
\label{eq23}     
\bm{P} = \bm{e}^s P_s = \partial L/\partial\bm{V} =  m(\bm{V} + \bm{u}),
\end{equation}
where the notation $\partial L/\partial\bm{V}$ means the gradient of $L$ w.r. to $\bm{V}$. The Euler-Lagrange equations in component form are $dP_s/dt = \partial L/\partial X^s.$ Expressing these equations using eqs.~(\ref{eq21}),~(\ref{eq22}), and~(\ref{eqA8}) yields the component form of the classical motion equations:
\begin{equation}
\label{eq24}     
m(g_{sq}\ddot{X}^q + [s,pr]\dot{X}^p\dot{X}^r) = -\partial_s W - m\partial u_s/\partial t + f_{sp}\dot{X}^p,
\end{equation}
where
\begin{equation}
\label{eq25}     
f_{sp} = m(\partial_s u_p - \partial_p u_s)
\end{equation}
are the covariant components of an antisymmetric rank two tensor, and [s,pr] are the Christoffel symbols of the first kind, defined by eqs.~(\ref{eqA11}) and~(\ref{eqA12}). Similarly, the Euler-Lagrange equations in $N$-vector notation are $d\bm{P}/dt = \bm{\nabla} L$, which yields
\begin{equation}
\label{eq26}     
md\bm{V}/dt = -\bm{\nabla}W -m\partial\bm{u}/\partial t + {\sf \bm{f}}\bm{\cdot}\bm{V}
\end{equation}
where ${\sf \bm{f}} = \bm{e}^s\bm{e}^p f_{sp}$ is the dyadic form of the rank two antisymmetric tensor mentioned above. The motion equation in this form looks like the usual Newton's 2nd law for Cartesian coordinates in Euclidean 3-space, but it applies to an arbitrary metric $N$-space. [The proof that eqs.~(\ref{eq26}) and~(\ref{eq24}) are equivalent is not quite trivial. It depends on using eqs.~(\ref{eq19}) and~(\ref{eqA7}), and recognizing that here the basis vectors are functions of the time-dependent coordinates $X(t)$, so that $\partial\bm{e}_p/\partial t = 0$,  but $d\bm{e}_p/dt = \dot{X}^q\partial_q\bm{e}_p$, whereby $d/dt = \partial/\partial t + \bm{V}\bm{\cdot}\bm{\nabla}$, the convective derivative, when acting on any vector field such as $\bm{V}$ or $\bm{u}$.]

The Hamiltonian for this general system is given by
\begin{equation}
\label{eq27}     
H = \bm{V}\bm{\cdot}\bm{P} - L = \frac{1}{2m}(\bm{P} - m\bm{u})^2 + W,
\end{equation}
where the last equality follows from eq.~(\ref{eq23}). Inserting components and basis vectors yields
\begin{equation}
\label{eq28}     
H = \frac{1}{2m}\bm{e}^p(P_p - mu_p)\bm{\cdot}\bm{e}^q(P_q - mu_q) + W
\end{equation}
which is the form that must be used in the transition to a statistical wave equation in which $P_p$ becomes a derivative operator that in general does not commute with either $\bm{e}^q$ or $u_q$, as we shall show in the next subsection. For the classical case treated here, eq.~(\ref{eq28}) reduces to the classical Hamiltonian $H_c$, given by
\begin{equation}
\label{eq29}     
H_c = \frac{1}{2m}g^{pq}(P_p - mu_p)(P_q - mu_q) + W,
\end{equation}
which is exactly what results if one starts with the usual definition $H_c = \dot{X}^p P_p - L$. Hamilton's canonical motion equations are simply $\dot{X}^p = \partial H_c/\partial P_p$, and $\dot{P}_p = -\partial H_c/\partial X^p$, which combine to yield eq.~(\ref{eq24}); or $\bm{V} = \partial H_c/\partial\bm{P}$ and $\dot{\bm{P}} = -\bm{\nabla}H_c$, which combine to yield the equivalent eq.~(\ref{eq26}).

     Perhaps the most important result of this subsection is the proof that all the standard equations and relations in the usual nonrelativistic classical mechanics notation in which three-vectors are written in boldface, including Newton’s second law, the Lagrangian, the conjugate momenta, the Hamiltonian, the Euler-Lagrange equations, and Hamilton's canonical equations, can be expressed in exactly the same boldface form in an $N$-dimensional metric configuration space of generalized curvilinear coordinates. Of course, in applying the equations, one must realize that each vector written in boldface has $N$ contravariant components that are the orthogonal projections of the vector onto the contravariant basis vectors, (and similarly for the $N$ covariant components), and that the metric and the affine connections must be specified in order to obtain the detailed $N$-space equations. We shall do examples later in this paper and in the next paper in this series to illustrate how one obtains and uses the basis vectors, the metric, and the connections. 

\subsection{Derivation of the Schr\"odinger equation in $N$ generalized coordinates}
\label{genSEQ}
     For the same reasons discussed in section\,\ref{SEMP}, the classical motions of particles of very small mass should be treated statistically. Assuming that coordinate trajectories $x^p = X^p(t)$ exist, then the fine-grained coordinate probability density is given by
           $$\rho^{f}(x,t) = |g_\cdot|^{-1/2}\prod_{q=1}^N\delta(x^q - X^q(t))$$ 
where $|g_\cdot|$ is the magnitude of the determinant of the covariant metric matrix (see the Appendix). The corresponding fine-grained probability current density is $\bm{j}^f = \bm{e}_p\dot{X}^p\rho^{f}(x,t).$ Using relations in the Appendix, it is easy to show that these quantities satisfy the $N$-space continuity equation $\partial_t\rho^{f}(x,t) + \bm{\nabla}\bm{\cdot}\bm{j}^f = 0$. The ensemble-averaged (smoothed) densities $\rho$ and $\bm{j}$ must also satisfy the continuity equation
\begin{equation}
\label{eq30}     
\partial_t\rho + \bm{\nabla}\bm{\cdot}\bm{j} = 0.
\end{equation}
This equation looks exactly the same as eq.\,(\ref{eq2}) for Euclidean 3-space. But here, the gradient operator $\bm{\nabla} = \bm{e}^p\partial_p$ is a sum of $N$ terms, as is $\bm{j}$, which is given by either $\bm{e}_p j^p$ or $\bm{e}^p j_p$; $x$ stands for the $N$ real independent variables (curvilinear coordinates) $x^1, \ldots ,x^N$; and the contravariant and covariant basis vectors $\bm{e}^p$ and $\bm{e}_p$ depend on the coordinates and are not unit vectors in general. 

     Now we may proceed just as in section~\ref{SEMP}. That is, {\em eqs.\,(\ref{eq3})-(\ref{eq10}) apply to the $N$-space system with no change in notation or form!}  We start with eq.\,(\ref{eq3}), $\bm{j}(x,t)=\rho(x,t)\bm{v}(x,t)$, and proceed step by step through eqs.\,(\ref{eq4})-(\ref{eq9}) until we get to eq.\,(\ref{eq10}), which we reproduce here:
\begin{equation}
\label{eq31}
 i\hbar\partial_t\psi = \frac{1}{2m}(-i\hbar\bm{\nabla} - m\bm{u})^2\psi + W\psi.
\end{equation}
Note that we have already chosen $\Gamma = \hbar$, in accordance with our discussion in Sec.~\ref{SEMP}. However, in eq.\,(\ref{eq31}), $\bm{\nabla}$ is the $N$-space gradient operator,  as noted just above; the unknown $N$-vector field $\bm{u}$ has $N$ covariant and $N$ contravariant components; and the wavefunction $\psi$, the unknown scalar field $W$, each of the components $u^p$, $u_p$, and each of the basis vectors may depend on all or some of the $N$ independent variables $x$. 

     The form-invariance of eq.\,(\ref{eq31}) when written in boldface vector notation should not be a great surprise: We use it in three dimensions every day, e.g., when we express the electric monopole SEQ in spherical polar coordinates $x=(r,\theta,\phi)$ for the unperturbed case $\bm{u} = 0$ and $W = W(r)$.
     
     For convenience in what follows, we write the analog of eq.\,(\ref{eq14}):
\begin{equation}
\label{eq32}
\bm{v}^{op} = m^{-1}\bm{p}^{op} - \bm{u};\;\;\;\;\bm{p}^{op} = -i\Gamma\bm{\nabla} = -i\hbar\bm{e}^p\partial_p,
\end{equation}
and we also define the operator
\begin{equation}
\label{eq33}
H^{op} = \frac{1}{2m}(\bm{p}^{op} - m\bm{u})^2 + W.
\end{equation}
Note that the presence of the combination $(\bm{p}^{op} - m\bm{u})$ ensures that the $N$-dimensional vector field $\bm{u}$ must have the same significance as the electromagnetic vector potential for a single electric monopole, that of a gauge field. In accordance with eq.\,(\ref{eq28}), $H^{op}$ must be written in terms of basis vectors and components as
\begin{equation}
\label{eq34}
H^{op} = \frac{1}{2m}\bm{e}^p(-i\hbar\partial_p - mu_p)\bm{\cdot}\bm{e}^q(-i\hbar\partial_q - mu_q) + W.
\end{equation}
Then the N-space statistical wave equation (\ref{eq31}) has the canonical form
\begin{equation}
\label{eq35}
i\hbar\partial_t\psi = H^{op}\psi.
\end{equation}
At this point, for clarity we write out eq.\,(\ref{eq34}):
\begin{equation}
\label{eq36}
H^{op} = \frac{1}{2m}[g^{pq}(-i\hbar\partial_p - mu_p) - i\hbar\bm{e}^p\bm{\cdot}\partial_p\bm{e}^q](-i\hbar\partial_q - mu_q) + W.
\end{equation}
Note that $H^{op}$ will include not only the derivatives $\partial_p u^q$, but also the term involving $\bm{e}^p\bm{\cdot}\partial_p\bm{e}^q$. As discussed in the Appendix, in conventional tensor calculus the first derivatives of the basis vectors can be written as linear combinations (LC's) of the basis vectors, and the coefficients of the LC's are the affine connections $\Gamma^{q}_{pr}$; see eq.~(\ref{eqA10}). If these connections are symmetric in their lower indices, then they are equal to the corresponding Christoffel symbols, which depend only on the metric and its first derivatives. This symmetry occurs {\it e.g.} in a global coordinate transformation from say Cartesian to curvilinear coordinates such as spherical polar coordinates in Euclidean 3-space. However, if some or all of the curvilinear coordinate basis vectors are referred to a locally Cartesian (or pseudo-Cartesian) set of basis vectors, which may always be done, then the resulting connections may be asymmetric in their lower indices, and one obtains a different expression for $\bm{e}^p\bm{\cdot}\partial_p\bm{e}^q$ than the conventional expression involving the Christoffel symbols. See eqs.~(\ref{eqA18})-(\ref{eqA21}). This kind of connection occurs in the treatment of particles with spin, in the next paper in this series.

     In order to identify $\bm{u}$ and $W$ in eq.\,(\ref{eq34}), we follow the same procedure used in section\,\ref{SEMP}, namely, the Ehrenfest theorem in reverse. First, we define the mean $N$-velocity vector of the system by 
\begin{equation}
\label{eq37}
\overline{\bm{V}}(t) = \int d^{N}x\,\sqrt{|g|}\,\bm{j}(x,t),
\end{equation}
the analog of the first part of eq.\,(\ref{eq11}). Then we substitute eq.\,(\ref{eq7}) with $\Gamma = \hbar$, valid here as an $N$-space equation, integrate by parts, and use eq.\,(\ref{eq32}) to obtain
\begin{equation}
\label{eq38}
\overline{\bm{V}}(t) = \int d^{N}x\,\sqrt{|g|}\,\psi^*\bm{v}^{op}\psi.
\end{equation}
In doing this integration by parts, we insist that i) all integrands that are bilinear in $\psi^*$ and $\psi$ must obey periodic boundary conditions in all coordinates, and ii) eq.~(\ref{eqA18}) must be valid. Just as in the case of the axiomatic SEQ, these requirements make the operators $\bm{v}^{op},\,\bm{p}^{op}$, and $H^{op}$ Hermitian. Now, we take the time derivative of the mean velocity. From eqs.\,(\ref{eq38}),\,(\ref{eq35}),\,(\ref{eq34}), and the Hermiticity of $H^{op}$, we obtain immediately
\begin{equation}
\label{eq39}
d\overline{\bm{V}}(t)/dt = \frac{1}{i\hbar}\int d^{N}x\,\sqrt{|g|}\,\psi^*[\bm{v}^{op},H^{op}]\psi.
\end{equation}
where $[\bm{v}^{op},H^{op}]$ is the commutator. The force operator $\bm{F}^{op}$ is 
\begin{equation}
\label{eq40}
\frac{m}{i\hbar}[\bm{v}^{op},H^{op}] \equiv \bm{F}^{op} = -\bm{\nabla}W - m\partial_t\bm{u} + \frac{1}{2}({\sf \bm{f}}\bm{\cdot}\bm{v}^{op} + \bm{v}^{op}\bm{\cdot}{\sf \bm{f}}),
\end{equation}
where
\begin{equation}
\label{eq41}
{\sf \bm{f}} = \bm{e}^p\bm{e}^q m(\partial_p u_q - \partial_q u_p).
\end{equation}
Equation\,(\ref{eq40}), valid in the $N$-space, is the analog of eq.\,(\ref{eq13}), valid in Euclidean 3-space. Note that the antisymmetric tensor term here replaces the cross-product term in eq.\,(\ref{eq13}); cross-products exist only in 3-spaces. Now,  comparing eqs.\,(\ref{eq40}) and (\ref{eq39}) with the generalized-coordinate Newton's second law, eq.\,(\ref{eq26}), it is immediately apparent that the Ehrenfest theorem is valid if and only if the $N$-space $H^{op}$ defined above is the usual classical system Hamiltonian with conjugate momentum $\bm{P}$ replaced by $\bm{p}^{op}$. That is, the analysis just above has identified the unknown functions $\bm{u}$ and $W$ that occur in the statistical wave equation constructed from the $N$-space probability continuity equation: The functions must be exactly those that appear in eq.\,(\ref{eq27}), the $N$-space classical Hamiltonian of the system. In addition, the overall analysis in this subsection has derived the general rule for “canonical quantization”, that the $N$-vector conjugate momentum $\bm{P}$ in any $N$-space classical Hamiltonian is to be replaced by $\bm{p}^{op} = -i\hbar\bm{\nabla} = -i\hbar\bm{e}^p\partial_p$, as given by eq.\,(\ref{eq32}). Note that this canonical quantization rule, $P_p\rightarrow p^{op}_p = -i\hbar\partial_p$, implies the usual commutator $[x^q , p^{op}_p] = i\hbar\delta^q_p$  in the $N$-space.          

\section{Examples}
\label{examples}
     In this section, we consider two simple examples that should help to clarify the generalized coordinate approach, namely, a system of two spinless point particles that may have diffeerent masses, and a system of arbitrarily many identical spinless point particles, in the unperturbed limit of two-body central force instantaneous internal interactions.  

\subsection{Two pointlike particles}
\label{2point}
     Let the masses be $m_1,m_2$, and choose three Cartesian coordinates for each particle's CM location, $x^1,x^2,x^3$ for particle 1, $x^4,x^5,x^6$ for particle 2. Then the kinetic energy $T$ on the trajectory ($x^p = X^p(t),\;p = 1, \ldots, 6$) is
\begin{equation}
\label{eq42}
T = \frac{1}{2}m[\frac{m_1}{m}\dot{X}^i\dot{X}^i +\frac{m_2}{m}\dot{X}^{i+3}\dot{X}^{i+3}] = \frac{1}{2}mg_{pq}\dot{X}^p\dot{X}^q,
\end{equation}
where the index $i$ ranges and sums from 1 to 3. From eq.~(\ref{eq42}), we may read off the diagonal 6-space metric: 
\begin{equation}
\label{eq43}
g_{11}=g_{22}=g_{33} = m_1/m;\;\;\;g_{44}=g_{55}=g_{66} = m_2/m,
\end{equation}
with other components zero. For this example, we consider the “unperturbed central force” case, by choosing $\bm{u} = 0$ and $W = W(r)_{x=X(t)}$ in the classical Lagrangian, where 
                                                 $$r = [(x^i - x^{i+3})(x^i - x^{i+3})]^{1/2}$$
is the distance between the particle CM's. The classical Hamiltonian is $H = T + W$, and $\bm{p}^{op} = -i\hbar\bm{\nabla} = -i\hbar\bm{e}^p\partial_p$, where the $\bm{e}^p$ are the Cartesian unit basis vectors $\bm{\hat{e}}^i$ and $\bm{\hat{e}}^{i+3}$. Thus, according to our general results in Sec.~\ref{schrodinger}, the Hamiltonian operator in the statistical SEQ is $H^{op} = (-\hbar^2/2m)\nabla^2 + W$, where $\nabla^2$ is given by eq. (A16). Since $g^{pq} = 1/g_{pq}$ for $p = q$ in this example, and zero otherwise, the SEQ is
\begin{equation}
\label{eq44}
i\hbar\partial_t\psi = -(\hbar^2/2m_1)\partial_i\partial_i\psi - (\hbar^2/2m_1)\partial_{i+3}\partial_{i+3}\psi + W\psi
\end{equation}
At this point, one may go to the conventional notation $x^i=x^i_1, x^{i+3}=x^i_2$, and then to CM and relative coordinates. 

     One reason for choosing this particular example is that it is probably the simplest two-particle example of the general method derived in section~\ref{schrodinger}. Another reason is to emphasize that what you get in the Hamiltonian operator in the derived SEQ is exactly what you have included in the classical Hamiltonian, no more and no less. For example, it is clearly physically incorrect to choose $\bm{u}$ = 0 and thus omit all incident and self radiation fields. It is also incorrect in principle to neglect retardation in two-body interactions, but that will be a negligible effect in cases involving slow motions of particles that remain close together. (Note that retardation for slowly-moving particles is not just a tiny relativistic correction if the particles are far apart.)

\subsection{Many pointlike particles}
\label{Npoint}
     Consider the extension of the two-particle system above to $N_p$ point particles, interacting with each other via two-body central force potential energies and also allowing external electromagnetic fields. We let the particles be identical, each with electric charge $q$, mass $m$, and CM location. Then the classical nonrelativistic Lagrangian, Hamiltonian, and motion equations each involve $N=3N_p$ coordinates, $x^p=X^p(t),\;p={1, \ldots, N}$. The development in Sec.~\ref{schrodinger} yields the general Schr\"odinger equation~(\ref{eq31}), or, equivalently, eqs.~(\ref{eq35}) and (\ref{eq36}), involving these $N$ coordinates. For this example, the metric may be chosen as the $N$-space Kronecker delta metric, $g_{pq}=\delta_{pq}=g^{pq}$, whereby the affine connections are zero, corresponding to three independent Cartesian coordinates for each particle. In order to achieve a familiar notation, we relabel the coordinates by letting $(x^p,\,p=1,N)\rightarrow(x_n^i,\,n=1,N_p,\,i=1,3)$, so that $n$ is a particle index and $i$ is a Cartesian coordinate index. Then, by analogy with the previous example, the simplest nontrivial “unperturbed” classical Hamiltonian contains $\bm{u} = 0$ and
\begin{equation}
\label{eq45}
W(x,t) = W^{int} = \frac{1}{2}\sum_{n=1}^{N_p}\sum_{n'=1}^{N_p}V(r_{nn'}),
\end{equation}
where terms with $n=n'$ are omitted from the double sum, $r_{nn'} = [(x_n^i - x_n'^i)(x_n^i - x_n'^i)]^{1/2}$, and $V$ is the two-body interaction energy that could involve not only the Coulomb repulsion but also other forces such as Yukawa interactions and gravity. If we allow given external electromagnetic potentials $\varphi^{ext}, A_i^{ext}$ to perturb the system, then the Hamiltonian would include the terms
\begin{equation}
\label{eq46}
W(x,t) = W^{int} + q\sum_{n=1}^{N_p}\varphi^{ext}(x_n,t);\;\;u_{n,i}(x,t) = (q/mc)A_i^{ext}(x_n,t),
\end{equation}
where $x_n$  stands for $(x_n^1,x_n^2,x_n^3)$. The $N$-vector $\bm{u} = \bm{e}^p u_p = \bm{e}_n^i u_{n,i}$, where $\bm{e}_n^i=\hat{\bm{e}}^i$, the Cartesian unit basis vector, the same for all $n$. Again we emphasize that the functions $W$ and $\bm{u}$ that appear in the Hamiltonian operator are exactly those that are chosen for inclusion in the classical Lagrangian and Hamiltonian. Clearly this often-used example once again neglects retardation and self-fields.

     It is important to note that for the identical particles in this example the total Hamiltonian is invariant under all pair interchanges of particle indices. This invariance leads immediately to the result that the total wavefunction solution of the general many-particle Schr\"odinger equation must either change sign under each pair interchange, or not change sign. As we know, the choice of sign change yields Fermions and the Pauli exclusion principle, while the choice of no sign change yields Bosons and Bose-Einstein condensation. Suppose one makes the sign change choice. Then, as discussed in detail by Schweber~\cite{Schweber}, the set of all Schr\"odinger equations for (1,2,3,...) identical particles is equivalent to the ``second quantized" many-particle field theory for Fermions in occupation number space. Likewise, if one makes the choice of no sign change, then the set of all Schr\"odinger equations for different numbers of particles is equivalent to the second quantized theory for Bosons. These equivalences combined with the results herein seem to imply that the conventional many-body quantum theories are mathematically compatible with particle trajectories, and are actually statistical theories.

\section{Summary, discussion, and prognosis}
\label{summary}
     {\bf Summary.} In sections~\ref{SEMP} and~\ref{schrodinger}, we showed the following: Given any system that requires $N$ independent real-valued curvilinear coordinates to describe classically. Given that the classical nonrelativistic kinetic energy of the system is bilinear in the coordinate first derivatives and thus defines a metric $N$-space. Also given that the force fields in the classical Lagrangian include non-negligible stochastic fields, so that the system must be described statistically. Then a correct statistical description of the system is always the usual $N$-coordinate Schr\"odinger equation, with an unknown constant $\Gamma$ in place of $\hbar$, in which the Hamiltonian operator is the classical Hamiltonian with the conjugate momenta replaced by momentum operators. The general $N$-vector momentum operator is simply $-i\Gamma$ times the $N$-space gradient operator. If the dominant stochastic field is the stochastic zero-point field (SZPF), then $\Gamma=\hbar$. In boldface $N$-vector notation, the resulting general $N$-coordinate nonrelativistic statistical Schrödinger equation is given by eq.~(\ref{eq31}),
$$ i\hbar\partial_t\psi = \frac{1}{2m}(-i\hbar\bm{\nabla} - m\bm{u})^2\psi + W\psi,$$
where $\bm{u}(x,t)$ is the $N$-vector velocity gauge field and $W(x,t)$ is the $N$-scalar potential energy that appear in the classical system Hamiltonian, $x = (x^1, \ldots, x^N)$  are the $N$ independent real curvilinear coordinates, $\bm{\nabla}=\bm{e}^q\partial_q$  is the $N$-space gradient operator, with $\bm{e}^q$ the contravariant basis vector normal to the hypersurface $x^q = const.^q$, and $m$ is a parameter having dimension mass that may be chosen at will, because the $N$-space metric contains the physical parameters. Furthermore, the multi-coordinate wavefunction in this statistical Schr\"odinger equation must satisfy exactly the same conditions as does the wavefunction in the usual axiomatic Schr\"odinger equation, whereby all wavefunction solutions are the same as the usual ones. We obtained these results by applying the reverse Madelung transform to the continuity equation for smooth coordinate probability density and flux, by using the covariant and contravariant basis vector approach to general tensor calculus in the metric $N$-space (see the Appendix), by applying the Ehrenfest theorem in reverse or by comparison with the classical Hamilton-Jacobi equation, and by comparison with well-known results for a classical charged radiation-damped harmonic oscillator immersed in the SZPF.

     In section~\ref{examples}, we treated two examples. The first was essentially trivial, a system of two point spinless particles having different masses in the unperturbed limit of an internal static central force interaction but no external fields. This example was included mainly to illustrate how the classical Hamiltonian and the corresponding operator reduce from their six-dimensional metric space expressions to the usual CM and relative coordinate expressions. The second example was a system of arbitrarily many identical spinless point particles in the limit of two-body central force non-retarded internal interactions, also in the presence of applied external fields. This example was included mainly to show how the $3N_p$-space Hamiltonian operator reduces to the usual form involving three Cartesian components of the CM position vector for each of $N_p$ particles, and also to emphasize that the set of derived Schrödinger equations for $N_p = 1,2,3,...$ predicts either fermions or bosons, as expected, and is equivalent to the usual quantized field description in occupation number space~\cite{Schweber}.  

     {\bf Discussion.} Since we have already discussed the mathematical approaches used in this work fairly thoroughly, in this subsection we will focus on interpretational and philosophical aspects of the results. In particular, we consider two important items, namely, whether classical trajectories are compatible with the conventional {\it interpretation} of the Schr\"odinger equation, and what could be the physical source of the wavelike properties of interference, diffraction, and tunneling predicted by the equation.

     The first item, discussed briefly in section~\ref{SEMP}, is that the statistical Schr\"odinger equation itself is mathematically compatible with classical coordinate trajectories. This compatibility arises because there are two levels of statistics involved in deriving the equation: First, one obtains a continuity equation for smooth coordinate probability density and flux by averaging the fine-grained density and flux, which contains Dirac deltas, over the random variables in the stochastic zero-point field. Then, using the methods discussed above, one obtains the statistical Schrödinger equation as an ensemble-averaged equation involving a  wavefunction that satisfies the usual conditions and is known {\it a priori} to have the usual statistical significance. 

     However, the conventional interpretation of discrete sets of energy eigenvalues as strictly quantized energies, the only energies allowed to the system, is not compatible with underlying classical trajectories. This incompatibility was discussed thoroughly in the papers by the author on the charged HO in SED~\cite{Goedecke1983}. In those papers it was shown that the HO energy is not sharp, even in the ground state, and furthermore that the system can never be in just a single excited state, despite the fact that the derived Schrödinger equation, which contains a radiation reaction vector potential that is not included in the conventional Schrödinger equation, yields the correct QED results for absorption and stimulated and spontaneous emission of electric dipole radiation, without  electromagnetic field quantization. Even in the ground state, direct calculation of ensemble averages from the detailed trajectory of the HO in the SZPF revealed that $\left<E^2\right>= 2\left<E\right>^2$, characteristic {\it e.g.} of Gaussian random variables, and denying strict quantization. That is, energy fluctuations must be present if underlying classical trajectories exist.

     It may be that this apparent interpretational dilemma can be resolved quite easily, as follows. We note again that the statistical Schrödinger equation derived in this work is an ensemble-averaged statistical equation. Therefore its predicted quantized eigenvalues of the Hamiltonian operator for an unperturbed bound system, the same eigenvalues predicted by the axiomatic SEQ, are {\it ensemble-averaged} energies $\left<E\right>$, which may form a discrete spectrum with no internal contradictions. If a known radiation field is applied to the system, then standard time-dependent perturbation theory applied to the statistical SEQ still yields the Einstein rule for line spectra, that emitted and absorbed angular frequencies are given by $\Delta\left<E\right>/\hbar$, and also yields the usual transition rules and (ensemble-averaged) transition probabilities per unit time. (As mentioned above, spontaneous emission results without quantized applied radiation fields if an appropriate RR vector potential is included in the classical Hamiltonian and thus in the SEQ Hamiltonian operator, but results only from the quantized applied radiation field if the RR potential is omitted.) The actual energies may fluctuate around the ensemble average energy eigenvalues without changing these results. However, we would expect the ensemble-averaged coordinate probability distributions predicted by the statistical Schr\"odinger equation to be enormously more probable than any others. We defer detailed consideration of energy fluctuations until later work.

    The second item mentioned above is the possible source of the wavelike properties of interference, diffraction, and tunneling predicted by the Schrödinger equation. The direct source must be the stochastic zero-point field (SZPF). In order to discuss how the SZPF could produce interference effects, let’s focus first on Casimir forces. These well-known “vacuum forces”, such as the force between two very large parallel plates separated by a small distance, or the force between two polarizable electric dipoles (the Casimir-Polder force) result from modification of the zero-point field induced by the presence of matter with which the field interacts. The standard results may be derived using either the quantum vacuum field or the SZPF, as discussed by de la Pe\~{n}a and Cetto~\cite{SED1}, and as verified in unpublished calculations by the author, ca. 1983. When using the SZPF, the Casimir-Polder force results from the ensemble average of the forces due to coherent multiple scattering of each SZPF plane wave mode and the consequent coherent zero-point oscillations of the two induced dipoles. The Casimir force between two parallel conducting plates also results from coherent multiple scattering between the plates, which alters the mode structure of the vacuum SZPF in the region between the plates and causes a net attractive ensemble average force between the plates. In both cases, coherent multiple scattering of the SZPF plane wave modes produces the interference and the "vacuum" forces. 

     Consider now a speculation on the famous problem of multiple-slit particle diffraction. It seems evident that the amplitudes and phases of the SZPF modes are altered from the free-field values by the presence of the matter in the plates containing the slits. These modes will be  different in the cases of no slits, one slit, two slits, etc. in a plane. Therefore, the forces on an approaching particle due to the SZPF must be different in each case, and perhaps it is not so surprising that the distribution of the transmitted particles in a beam is quite different in each case. The Schr\"odinger/statistical wave equation derived herein predicts interference patterns  as if each incident particle were a wave having a given ensemble-averaged incident direction and wavelength equal to Planck's constant divided by the incident momentum. Of course a single particle actually makes a dot on the detector screen behind the slits. But the interference pattern is indeed reproduced after a very large number of identical non-interacting particles pass through the slits, whether they pass through one at a time or all in a bunch.

     With respect to tunneling through a potential barrier, a particle following a classical trajectory can do this only if temporarily it receives enough energy to be kicked over the barrier. Such temporary or "virtual" energy transfers will occur in the presence of the SZPF, and their statistics must be described by the Schr\"odinger/statistical wave equation.  

     Let us add a comment about our identification of the unknown constant in the statistical wave equation as Planck's constant. Our comparison with the results for the charged harmonic oscillator in SED is equivalent to  comparison with experiment or with the axiomatic Schr\"odinger equation, except for our argument involving the SZPF and its close relation to the quantum vacuum field, which suggests the universality of the constant. A derivation of the numerical value of the constant is lacking, as has always been the case. For that matter, to the best of our knowledge there are no accepted derivations of the values of any of the fundamental constants. One expects that their values, which Dirac suggested might be time-dependent, are determined by the history and structure of the whole universe. In particular, as a classical albeit stochastic field, the SZPF must have sources, i.e., it must originate from (zero-point) oscillations of all the particles in the universe. Therefore there must be some self-consistency requirements involved that would determine values of one or more of the fundamental “constants”. 

Another point should be made about the universality of the unknown constant $\Gamma$ in the derived statistical Schr\"odinger wave equation: It seems clear that it should have the same value, $\hbar$, for every electrically charged fundamental particle that interacts via the Lorentz force with the SZPF. Furthermore, most  fundamental particles are themselves charged or are composed of charged particles. Because of the extremely intense ultra-high-frequency amplitudes in the SZPF, which has a power spectrum $\propto\omega_k^3$, the SZPF should interact via the Lorentz force with each electrically charged constituent of any particle, {\it e.g.} with the quarks in a neutron. However, neutrinos are outstanding counterexamples. Should their statistical quantum equations also have $\Gamma=\hbar$? We defer consideration of this question to later work.

     {\bf Prognosis}. In the next paper in this series, we utilize the results of this paper, in particular eqs.~(\ref{eq35}) and~(\ref{eq36}), to show that a straightforward statistical representation of the nonrelativistic rigid rotations of a charged massive object in terms of Euler angle coordinates and principal moments of inertia yields exactly the properties of quantum spin and the quantum interaction of a magnetic dipole with a magnetic field, including odd-half-integer as well as integer spin. We show that odd-half-integer spin particles cannot access integer spin states, and vice-versa. We also provide one simple way to overcome the well-known objection that an extended rigidly rotating electron model must involve supraluminal speeds. 

     If the {\it a priori} statistical treatment developed in the first two papers in this series is to be viable, future work must include fully relativistic treatments of spinless and spinning particles utilizing the same approach. While we have made significant progress toward these treatments, as of this writing quite a bit remains to be done.

     Another point that we feel should be emphasized again: As implied by our results, quantum mechanics, axiomatic or statistical, is only as good as the classical mechanics that underlies it. When we set up a quantum mechanics problem, we first need to decide how many “classical” coordinates are needed. (For example, consider a diatomic molecule. Do we need only CM and (two) Euler angle coordinates, or should we include the “internal” vibration coordinate? It’s up to us. If we think ambient energies will be fairly large, then we need the latter.) Then we must decide how to represent the interaction potential energies and gauge fields in the classical Lagrangian/Hamiltonian. Usually we are forced to make approximations. (For the diatomic molecule, we often use the lowest order approximation by assuming only an instantaneous harmonic central force between the atoms. This is incorrect in principle, because it neglects retardation and vector potentials, including external and self radiation fields, but it is also clearly quite accurate in some applications.) Again, it's up to us. So the traditional approach in classical mechanics, to choose the appropriate coordinates and interactions and obtain the Lagrangian and Hamiltonian, must still be done before one does the (statistical) quantum mechanics of the problem.

     Finally, we offer a further comparison of axiomatic nonrelativistic quantum mechanics (QM) with the statistical nonrelativistic QM developed herein. First, we re-emphasize that the two produce identical algebraic or numerical results for any and all choices of the classical Hamiltonian, for any nonrelativistic system whatsoever. We must remember that on its way to becoming axiomatic the original Schr\"odinger equation (SEQ) was inferred, not derived, by seeking the simplest linear second-order homogeneous wave equation whose eikonal limit is the Hamilton-Jacobi equation for the classical system considered. The probabilistic significance of the assumed smooth wavefunction $\psi$ and density $\psi^*\psi$ had to be determined {\it a posteriori}, as did the appropriate boundary conditions on the wavefunction, by comparison with experiment or by postulate. The significance of discrete sets of eigenvalues of the Hamiltonian operator as quantized energies,  the only energies allowed, seemed obvious, but actually constituted another postulate. In contrast, the fully statistical theory is unavoidable. Except for identification of the unknown real constant $\Gamma$, the theory follows inevitably once one decides to treat the nonrelativistic classical motions of small particles statistically. It derives the general canonical quantization rule and SEQ for a smooth wavefunction with interpretations and boundary conditions known {\it a priori}; but it does not support quantized energies. (As mentioned above, the statistical SEQ is an ensemble-averaged equation, whereby its eigenvalues ought to be ensemble-averaged energies.) If we do insist on truly quantized energies, then we are choosing to maintain dual theories that seem to yield the same results. Therefore, it is most important that we continue to investigate the statistical approach in an effort to determine what experiments could distinguish between the two theories. 

\section{Acknowledgements}
\label{acknowledgements}
The author would like to thank Stephen Pate, Michael Engelhardt, and Stefan Zollner for helpful discussions and assistance in preparing the manuscript.

\appendix*
\section{Generalized coordinates and coordinate basis vectors}
\label{appendix}
To the best of our knowledge, only the well-known text by Lichnerowicz~\cite{Lichnerowicz} and a recent paper by the author~\cite{Goedecke2011} provide fairly complete discussions of the basis vector approach to tensor calculus in general metric spaces. This very clear, efficient, and versatile approach is particularly well-suited to the material treated in this work. In the interest of readability, we provide a brief summary below.

Consider a set of real continuous independent variables $x = (x^1, \ldots, x^N)$ as curvilinear coordinates in an $N$-dimensional manifold or ``$N$-space".  $N$ may be finite or infinite. Each coordinate $x^p,\;p = 1, \ldots, N$ may be non-compact or compact and may have any physical dimension. The $(N-1)$-dimensional coordinate hypersurfaces are defined as $[\Sigma_p:\;x^p = const.^p]$ , and the coordinate curves are given by $[C_p:\;x^q = const.^q,\,\forall{\,q\neq p}]$; {\it i.e.}, a particular $C_p$ that passes through a point $x$ is defined as the intersection of the $(N-1)$ hypersurfaces $[\Sigma_q,\,q\neq p]$ that contain $x$.  We may define a set of $N$ linearly independent (LI) vectors  $\bm{e}_p$, called ``covariant" basis vectors, as tangents to the $[C_p]$ passing through point $x$, and a so-called dual LI set $\bm{e}^p$, called ``contravariant" basis vectors, as normals to the $[\Sigma_p]$ at point $x$; either set will serve as a set of coordinate basis vectors for vector and tensor fields in the tangent linear vector space. The designations ``contravariant" and ``covariant" refer to how quantities transform under a general coordinate transformation in the $N$-space. For single-index quantities, ``contravariant", indicated by a superscript index, means that the $\bm{e}^p$ and other single-superscript quantities transform as do the coordinate differentials $dx^p$; ``covariant" means that the $\bm{e}_p$ and other single-subscript quantities transform as do the partial derivatives $\partial_p$. 

The inner (dot) products among the normal and tangent basis vectors at any point are given by
\begin{equation}
\label{eqA1}
\bm{e}_p\bm{\cdot}\bm{e}^q = \bm{e}^q\bm{\cdot}\bm{e}_p = \delta_p^q,
\end{equation}
the Kronecker delta. We may write any $N$-vector field $\bm{A}(x)$ as linear combinations (LC's) of the coordinate basis vectors:
\begin{equation}
\label{eqA2}
\bm{A}(x) = \bm{e}_p(x)A^p(x) = \bm{e}^q(x)A_q(x).
\end{equation}
We adopt the extended Einstein summation convention that repeated indices in any term of an equation, one "up" and one "down", or both up or both down, are summed over from 1 to $N$. The coefficients of the LC's are also the components of the vector, {\it i.e.}, $A^q = \bm{e}^q\bm{\cdot}\bm{A}$ are the contravariant components, and $A_q = \bm{e}_q\bm{\cdot}\bm{A}$ are the covariant components. Similarly, we may write any 2nd rank tensor or ``dyadic" field as bilinear combinations of the basis vectors; the coefficients of the bilinear combinations have two indices and are the ``components" of the dyadic. Also, any triadic or third rank tensor is a trilinear combination of basis vectors; etc. All such components are present in the conventional Riemann-Einstein version of tensor calculus. However, that version does not use basis vectors and thus cannot write the invariant quantities such as an $N$-vector field $\bm{A}$ in the general and very convenient boldface notation of eq.~(\ref{eqA2}). 

The displacement vector $d\bm{x}$ connecting two infinitesimally separated points and the gradient operator $\bm{\nabla}$ are written as the following LC’s:
\begin{equation}
\label{eqA3}
d\bm{x}   = \bm{e}_p dx^p;\;\;\;\bm{\nabla} = \bm{e}^p \partial_p,
\end{equation}
where $\partial_p = \partial/\partial x^p.$ As in relativity, $d\bm{x}$ is chosen to have dimension length ($L$) and $\bm{\nabla}$ to have dimension $1/L$. Therefore, in general, $\bm{e}^p$ and $\bm{e}_p$  may not be unit vectors, may not be dimensionless, may not have the same dimension, and may not even be parallel; and the covariant and contravariant components of a vector may not have the same dimension or the same dimension as the vector itself. 

The square of the line element $ds$ between infinitesimally separated points is defined conventionally by
\begin{equation}
\label{eqA7}
\pm ds^2 = d\bm{x}\bm{\cdot}d\bm{x} = \bm{e}_p\bm{\cdot}\bm{e}_q dx^p dx^q = g_{pq}(x)dx^p dx^q,
\end{equation}
where either + or - may be chosen, and the last equality defines the (covariant components of the) metric $g_{pq}$  and thereby the fundamental symmetric dot products among the subscripted basis vectors associated with these $N$ curvilinear coordinates. In general $ds^2$ may be positive, negative, or zero. The dot product of any basis vector with itself may be either positive or negative, yielding an “indefinite metric”, necessary here and in relativity since all coordinates are chosen to be real-valued. Using eqs.~(\ref{eqA1}) and~(\ref{eqA2}) yields
\begin{equation}
\label{eqA8}
\bm{e}_r\bm{\cdot}\bm{A} = A_r = g_{rp}A^p;\;\;\;\bm{e}^p\bm{\cdot}\bm{A} = A^p = g^{pq}A_q,
\end{equation}
where
\begin{equation}
\label{eqA9}
g^{pq} = \bm{e}^p\bm{\cdot}\bm{e}^q
\end{equation}
comprise the “contravariant components” of the metric, or in common usage, simply the contravariant metric. Also, eq.~(\ref{eqA8}) implies $g^{rp}g_{pq} = \delta^r_q$, i.e., the matrix $(g^{\cdot})$ with elements $(g^\cdot)_{pq} = g^{pq}$ is the inverse of the matrix $(g_{\cdot})$ with elements $(g_{\cdot})_{pq} = g_{pq}$. Note that in general indices may be raised and lowered using the appropriate components of the metric.

In Lichnerowicz' basis-vector approach to tensor calculus, the coordinate partial derivatives of the $N$-space basis vectors at point $x$ are expressible as LC's of those basis vectors: 
\begin{equation}
\label{eqA10}
\partial_p\bm{e}_q = \Gamma^r_{pq}\bm{e}_r;\;\;\;\partial_p\bm{e}^r = -\Gamma^r_{pq}\bm{e}^q.
\end{equation}
The $ \Gamma^r_{pq}$ are the so-called affine connections; here, we are following Hartle's~\cite{Hartle} choice for the ordering of the subscripts on these connections, opposite to that used in the author's recent paper~\cite{Goedecke2011}. (In the conventional Riemann-Einstein version of tensor calculus, the affine connections are defined by considering ``parallel transport" of vector components). A space in which the connections are symmetric in their lower indices is called “torsion-free”. Virtually all efforts in relativity have assumed this property. In such cases it is straightforward to show that
\begin{equation}
\label{eqA11}
\Gamma^r_{pq} = \{^r_{pq}\} = g^{rs}[s,pq],
\end{equation}
\begin{equation}
\label{eqA12}
[s,pq] =\mbox{\footnotesize$\frac{1}{2}$}[\partial_p g_{qs}+\partial_q g_{ps} - \partial_s g_{pq}],
\end{equation}
where $[s,pq] = [s,qp]$ is called a ``Christoffel symbol of the first kind", and $\{^r_{pq}\}$ is called a ``Christoffel symbol of the second kind". These symbols satisfy an important identity,
\begin{equation}
\label{eqA13}
\{^q_{pq}\} = \partial_p\ln\sqrt{|g_\cdot|},
\end{equation}
where $|g_\cdot|$ is the magnitude of the determinant of the matrix $(g_\cdot)$. Consider the gradient of a vector field,
\begin{equation}
\label{eqA14}
\bm{\nabla A} = \bm{e}^q \partial_q ( \bm{e}_p A^p) = \bm{e}^q\bm{e}_r(\partial_qA^r + \Gamma^r_{qp}A^p),
\end{equation}
where the last term follows from eq.~(\ref{eqA10}). The quantity $(\partial_qA^r + \Gamma^r_{qp}A^p)$, denoted $A^r_{\;\,;q}$ or $D_qA^r$ or $\nabla_qA^r$, is called the covariant derivative of the vector (component) $A^r$.  The open vector product $\bm{\nabla A}$ is called a “dyadic” (or rank two tensor); the quantities $D_qA^r = \bm{e}_q\bm{\cdot}\bm{\nabla A}\bm{\cdot}\bm{e}^r$ are the $N^2$ mixed  components with first index down (covariant), second index up (contravariant). The divergence $\bm{\nabla}\bm{\cdot}\bm{A}$ of a vector field is an important quantity. To obtain it, just put a dot between the basis vectors in eq.~(\ref{eqA14}), and use eqs.~(\ref{eqA1}) and~(\ref{eqA13}):
\begin{equation}
\label{eqA15}
\bm{\nabla}\bm{\cdot}\bm{A} = (\sqrt{|g_\cdot|})^{-1}\partial_q(\sqrt{|g_\cdot|}A^q).
\end{equation}
If $\bm{A}=\bm{\nabla}\Phi$, where $\Phi$ is any (scalar) function of $x$, then
\begin{equation}
\label{eqA16}
\bm{\nabla}\bm{\cdot}\bm{A} = \nabla^2\Phi = (\sqrt{|g_\cdot|})^{-1}\partial_q(\sqrt{|g_\cdot|}g^{qp}\partial_p\Phi).
\end{equation}
The N-space volume element $dV_N$ is also needed; it is
\begin{equation}
\label{eqA17}
dV_N = d^Nx\sqrt{|g_\cdot|},\;\;\;d^Nx = dx^1, \cdots, dx^N.
\end{equation}
An important identity is needed to show that various operators are Hermitian:
\begin{equation}
\label{eqA18}
\int d^Nx\sqrt{|g_\cdot|}\bm{\nabla}f = 0,
\end{equation}
where the integration extends over all $N$-space and $f$ is any function. In order to prove this identity, simply write $\bm{\nabla}=\bm{e}^p\partial_p$, integrate by parts, use eqs.~(\ref{eqA10}),(\ref{eqA11}), and~(\ref{eqA13}), and require that the integrand $\sqrt{|g_\cdot|}\bm{e}^q f$  satisfy periodic boundary conditions or go to zero at the coordinate boundaries.

     Up to this point, this appendix has considered conventional tensor calculus with coordinate basis vectors in a metric space of $N$ dimensions. The designation ``conventional" here implies that i) the affine connections are symmetric in their lower indices, and ii) eqs.~(\ref{eqA8}) and~(\ref{eqA10}) are valid. Together, these equations imply that the $N$-space considered is completely isolated from other possible coordinate spaces. For example, suppose that a total space consists of an $N$-space and an $M$-space. In order for these two subspaces to be isolated from each other, i) the metric must be globally transformable to block diagonal form, $N\times N \oplus M\times M$, as implied indirectly by eqs.~(\ref{eqA8}) and~(\ref{eqA10}), and ii) the derivatives of basis vectors in each subspace must be expressible as LC's of basis vectors only in that subspace, as implied by eq.~(\ref{eqA10}). 

     Non-isolated subspaces are actually quite common~\cite{Goedecke2011}; they should be important in a fully relativistic treatment. In this nonelativistic treatment, we work with isolated spaces. However, we do need to consider another deviation from conventional tensor calculus, namely, suppose the coordinate basis vectors $(\bm{e}_p,\bm{e}^q)$ are referred locally to a given Cartesian set $[\bm{\hat{e}}_i = \bm{\hat{e}}^i, \;\hat{g}_{ij} = \bm{\hat{e}}_i\bm{\cdot}\bm{\hat{e}}_j = \delta_{ij}]$, where the indices $i,j,k,...$ range and sum over the same set of integers as the indices $p,q,r,...$ (Just as in relativity, such a referral may always be done using local coordinate transformations.) That is, suppose that 
\begin{equation}
\label{eqA19}
\bm{e}_q = A_{iq}\hat{\bm{e}}_i;\;\;\;\bm{e}^q = A^{-1}_{qj}\hat{\bm{e}}_j,
\end{equation}
where the coefficients  are given functions of the curvilinear coordinates, valid in a neighborhood around any point $x$. Note that the second relation in eq.~(\ref{eqA19}) is implied by the first and eq.~(\ref{eqA1}). The metric is then given by 
\begin{equation}
\label{eqA20}
g_{pq} = \bm{e}_p\bm{\cdot}\bm{e}_q = A_{ip}A_{iq};\;\;\;g^{pq} = \bm{e}^p\bm{\cdot}\bm{e}^q = A^{-1}_{pj}A^{-1}_{qj}.
\end{equation}
The affine connections are then obtained by derivation of eq.~(\ref{eqA19}), using eq.~(\ref{eqA10}):
\begin{equation}
\label{eqA21}
\Gamma^q_{pr} = (\partial_pA^{-1}_{qj})A_{jr},
\end{equation}
keeping in mind that in a locally Cartesian system underlying the basis vectors $[\hat{\bm{e}}_i]$ at point $x$, the first derivatives of these vectors (and thus the locally Cartesian system connections) vanish at $x$. Then eqs.~(\ref{eqA10}), (\ref{eqA19}) and (\ref{eqA20}) yield
\begin{equation}
\label{eqA22}
\bm{e}^p\bm{\cdot}\partial_p\bm{e}^q = -g^{pr}\Gamma^q_{pr} = A^{-1}_{pi}\partial_pA^{-1}_{qi}.
\end{equation}
Note that these $\Gamma^q_{pr}$ may not be symmetric in their lower indices. Such seems to be the case for the 3-space of the Euler angles, as mentioned in Sec.~\ref{genSEQ} above and derived in the next paper in this series. Nevertheless, the identity~(\ref{eqA18}) is still valid in the Euler angle space, as shown in that paper.

\end{document}